\documentclass[mathleft
]{an}
\usepackage{times}
\usepackage{graphicx}
\usepackage{times}

\usepackage[british]{babel}

\begin{document}

% The following seven commands are intended for editorial usage and should be ignored by
% the author(s).
\Pagespan{789}{}% Document's page range.
% If second parameter is left empty, the last page is computed automatically.
\Yearpublication{2006}%
\Yearsubmission{2005}%
\Month{11}%
\Volume{999}%
\Issue{88}%
% \DOI{This.is/not.aDOI}%

\title{Results of the photometry of the spotted dM1-2e star EY Draconis}
\author{K. Vida \inst{1,2}}

\titlerunning{Photometry of EY Draconis}
\authorrunning{K. Vida, \dots}

\institute{
E\"otv\"os Lor\'and University, Budapest, Hungary
\and
Konkoly Observatory of the Hungarian Academy of Sciences, Budapest, Hungary}

\received{30 May 2005}
\accepted{11 Nov 2005}
\publonline{later}

\keywords{starspots -- stars: activity  -- stars: late-type -- stars:
imaging -- stars: individual (EY~Dra)}

\abstract{
 We have observed EY Draconis with the 24`` telescope of Konkoly Observatory in Budapest for 64 nights. In the first observing season the star produced a stable light curve for more than 60 rotation periods, however, the  light curves observed in the next season and the spot modelling show clear evidence of the evolution of the spotted stellar surface. The changes of the maximum brightness level suggests the existence of a longer period of about 300 days, which seems to be confirmed by the ROTSE archival data.
}

\maketitle
\sloppy
\section{Introduction}
EY Draconis was first categorised as an active star in 1991 when the ROSAT X-ray/EUV satellite did an all-sky survey, and classified the object named RE~1816+54 as an EUV source, which optical counterpart was identified as EY Draconis. \cite{Jeffries:MNRAS94} presented spectroscopic measurements, and classified the star as a fast-rotating dM1-2e type star with chromospheric and coronal activity, based on the observed H$\alpha$ and Ca H\&K emission lines. \cite{Eibe:AA98} studied the H$\alpha$ line, and detected the presence of plage areas, and a flare. \cite{Barnes:MNRAS01} presented Doppler imaging based on high-resolution spectroscopy, and derived stellar parameters as well as the image of the stellar surface, where starspots are found mainly at high latitudes.

The only photometry published up to now was made by \cite{Robb:IBVS95}. The Johnson V-band measurements covered 8 days. From the observed light curve -- which appeared to be stable through the time of the observations -- they concluded that the variations are caused by at least two large spots or spot groups. The authors found a period of 0.459 days in the light variability. The increased scatter of the phased $V$ light curve in the minima suggested the presence of a small flare.

Stellar parameters obtained from the mentioned articles are summarised in Table~\ref{tab:params}.

\begin{table}
%\centering
\begin{tabular}{lcl}

 \noalign{\smallskip}
Parameter&value&ref.\\
 \hline \hline
  \noalign{\smallskip}
  spectral type &dM1-2e& $^1$ \\
  $v\sin i$ & 61 km s$^{-1}$& $^1$\\
  $P_{\mathrm{rot}}$ & 0.459 d &$^2$\\
  distance & 45.5 $\pm$2.1 pc& $^3$\\
  $M_V$& 8.54 $\pm$ 0.12& $^3$\\
  $r\sin i$ & 0.549 $\pm$ 0.002 R$_\odot$& $^3$\\
  $i$ & ~70$^\circ$& $^3$\\
  \noalign{\smallskip}
  \hline
  \noalign{\smallskip}
  $^1$ \cite{Jeffries:MNRAS94}\\
  $^2$ \cite{Robb:IBVS95}\\
  $^3$ \cite{Barnes:MNRAS01}

\end{tabular}
\caption{Stellar parameters of EY Draconis}
\label{tab:params}
\end{table}

\section{Observations and data reduction}
Our measurements were made using the 60-cm telescope of the Konkoly Observatory in Budapest equipped with a Wright $750\times1100$ pixel CCD using $BV(RI)_C$ passbands. The field of view was $17'\times24'$. More than 2000 frames were taken in each passband on 64 nights. The data reduction and differential aperture photometry was performed using the standard IRAF\footnote{
IRAF is distributed by the National Optical Astronomy Observatory, which
is operated by the Association of Universities for Research in Astronomy, Inc.,
under cooperative agreement with the National Science Foundation.
} 
packages. We used GSC 03904-00259 as check, and GSC 03904-00645 as comparison star. Note, that the comparison and check stars are not the same as those in \cite{Robb:IBVS95}, since we found GSC 03904-00259 to be closer to EY Draconis in colour. The correction for the atmospheric extinction was based on the work of \cite{Hardie} using the star fields of EY Draconis, TZ Aurig\ae~ and SS Cancri. The effect of principal extinction coefficient is negligible in our case, because the size of the CCD chip is small, and all the stars are close to each other. The secondary extinction coefficient for $B-V$ colour index is $k_{bv}''=−0.027$. In this paper only $V$-band data are used.

\begin{figure}

\centering
\includegraphics[width=83mm,bb=50 50 410 302]{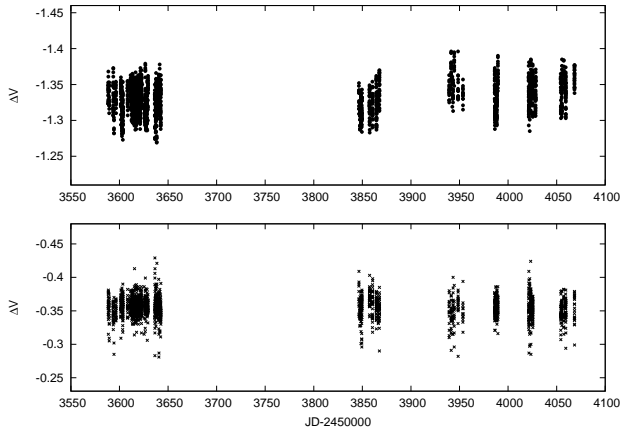}
\caption{EY Draconis (top) and the comparison -- check light curve (bottom) in $V$ colour, from the Budapest observations}
 \label{fig:all_v}
\end{figure}

The $V$ light curve can be seen on Figure \ref{fig:all_v} together with the comparison--check light curve. The scatter of the data seems to be the same on both plots. The reason of this is, that the comparison star is more than one magnitude fainter as EY Draconis, and the check star is  0.35 magnitude fainter than the comparison. This causes an increase in the errors of the measurements. So while the ''scattering'' of the EY Draconis light curve is mostly caused by the rotational modulation of the star as confirmed by the Fourier-analysis (see Section \ref{sect:analysis}), the scatter on the comparison--check light curve reflects the noise of the measurements.

Additional 365 day-long photometric data were collected from the \cite{ROTSE} database.

\begin{figure}
	\centering
	\includegraphics[width=83mm,bb=50 50 410 302]{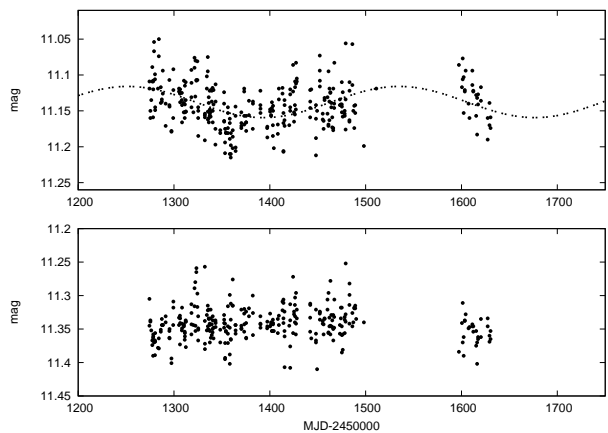}
	\caption{EY Draconis lightcurve obtained from the ROTSE database (top) and the difference between the lightcurves of two non-variable field star from the ROTSE database (bottom).}
	\label{fig:rotse}
\end{figure}

\section{Evolution of the stellar surface}
\begin{figure}
 \centering
 \includegraphics[width=83mm]{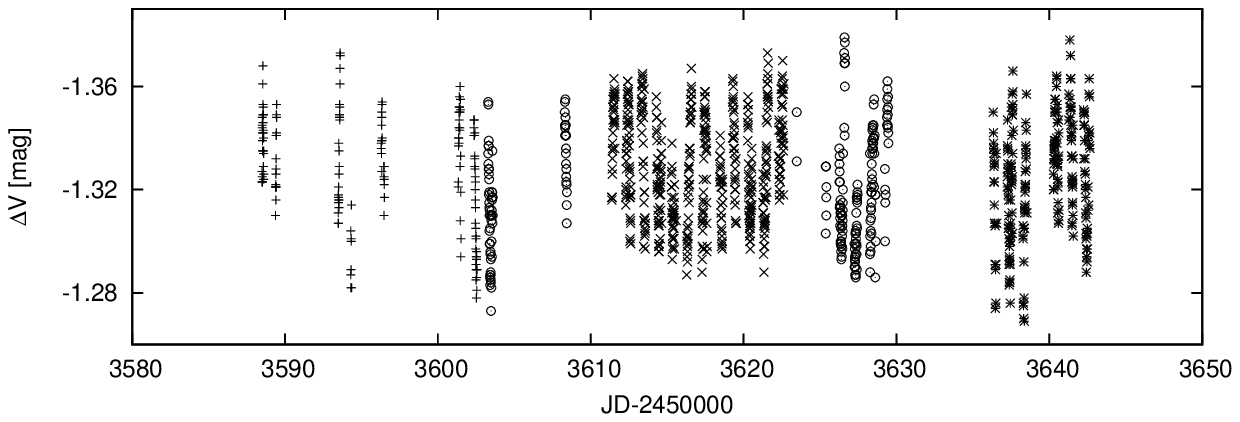}
 \includegraphics[width=67mm]{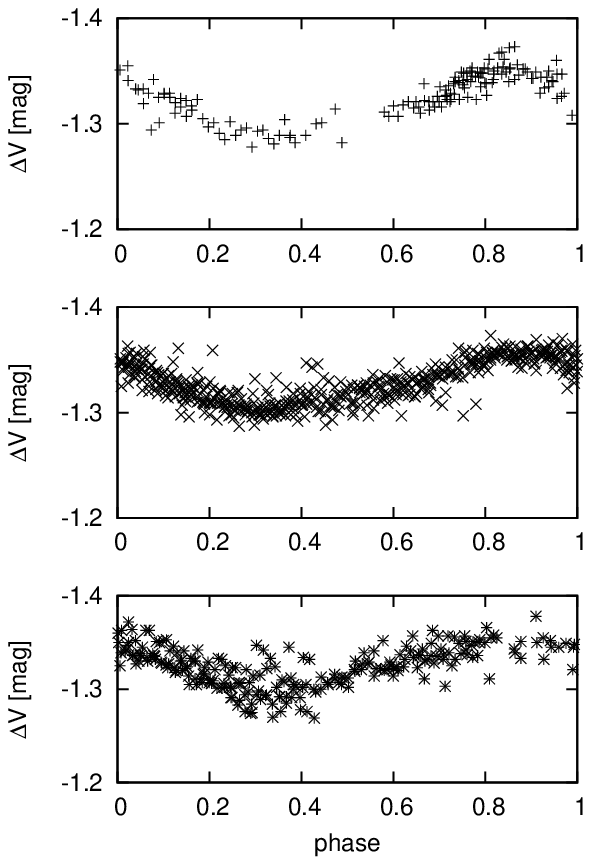}
 \caption{The measurements during the first observational season. On the top panel the obtained light curve can be seen, while on the lower three different phased parts of this light curve is plotted. The different symbols on the top light curve show the parts which were used to plot the phase--magnitude diagram.}
\label{fig:firstLC}
\end{figure}

The first, longest observational season covers 118 rotations, when the shape of the light curve remains practically unchanged (see Figure \ref{fig:firstLC}). This stability might be caused by an unresolved binary component, or large scale magnetic field, which could be a possible explanation for the stable prominence clouds  detected by \cite{Eibe:AA98}.

We used SML (SpotModeL) software made by \cite{Ribarik:AN03} for spot modelling of the light curves. This programme is able to model lightcurves with up to three circular spots, by varying their positions and sizes. The spot latitudes were chosen based on the Doppler-images of \cite{Barnes:MNRAS01}, where high-latitude and polar spots can be seen. These parameters were fixed during the fit, since the latitudes of the spots cannot be determined from photometric observations. The minima of the phased light curves were chosen as initial spot longitudes, and one polar spot was assumed. 

\begin{figure}
	\centering
	\includegraphics[width=83mm,bb=50 50 410 302]{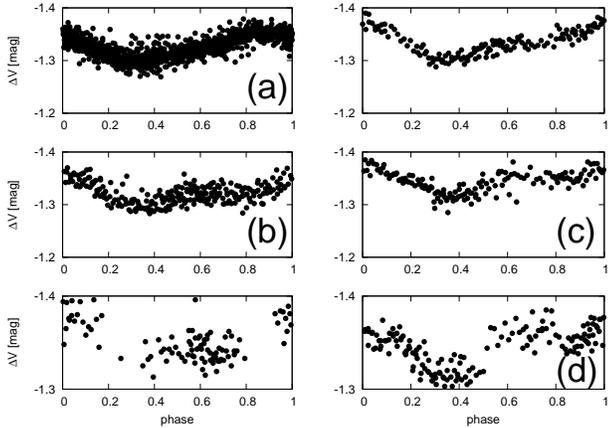}
	\caption{Phased light curves of EY Draconis. The plots show the magnitude change of the star in $V$ colour on the left side between 2400000+
53588-642 (a),
53846-67 (b),
and 53939-53; on the right side between
53986-89,
54021-25 (c) and
54054-68 (d) Julian dates.}
	\label{fig:phases}
\end{figure}

\begin{figure}
	\centering
	\includegraphics[width=41mm,bb=14 14 335 255]{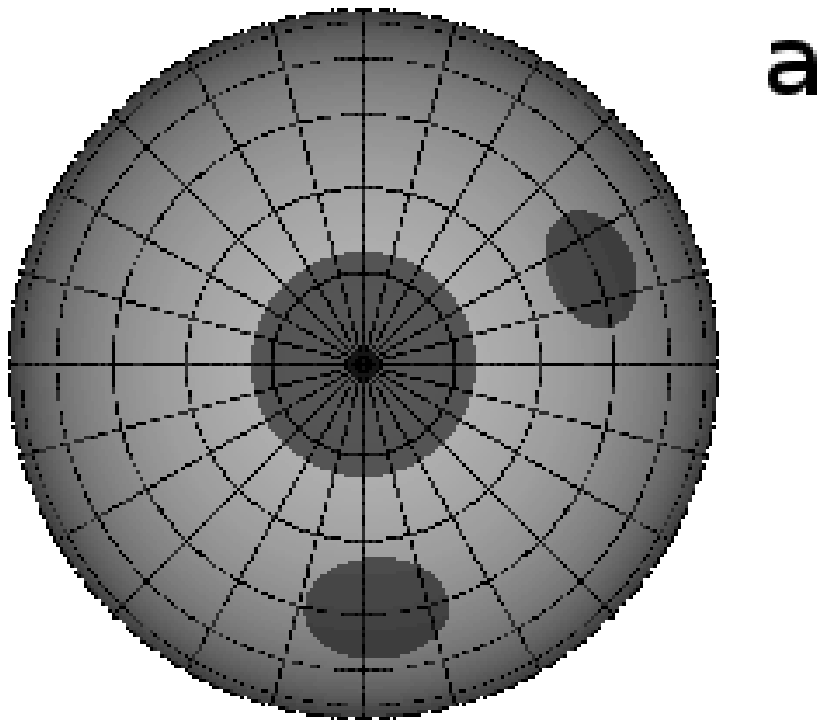}
	\includegraphics[width=41mm,bb=14 14 335 255]{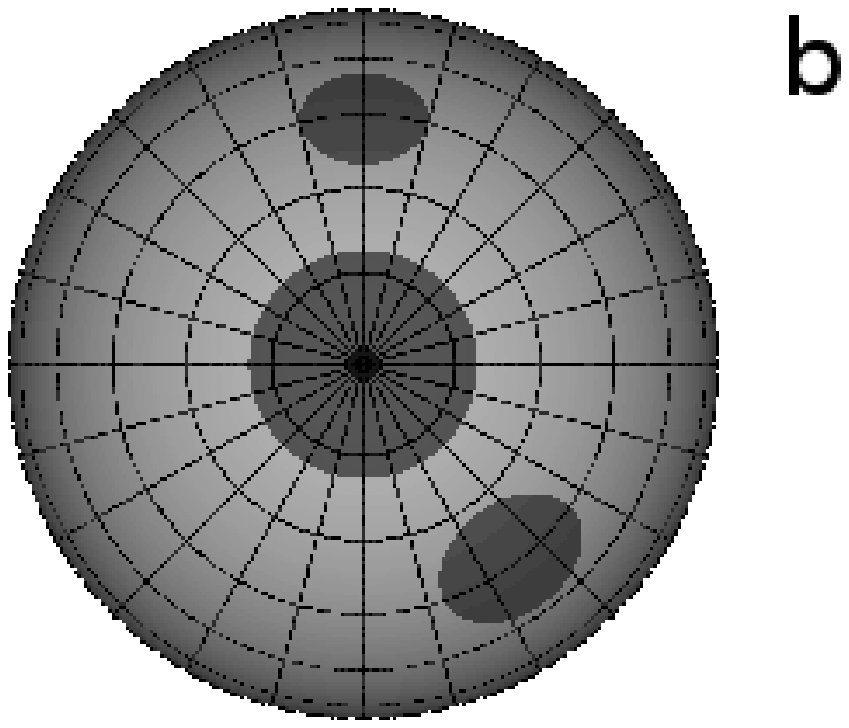}
	\includegraphics[width=41mm,bb=14 14 335 255]{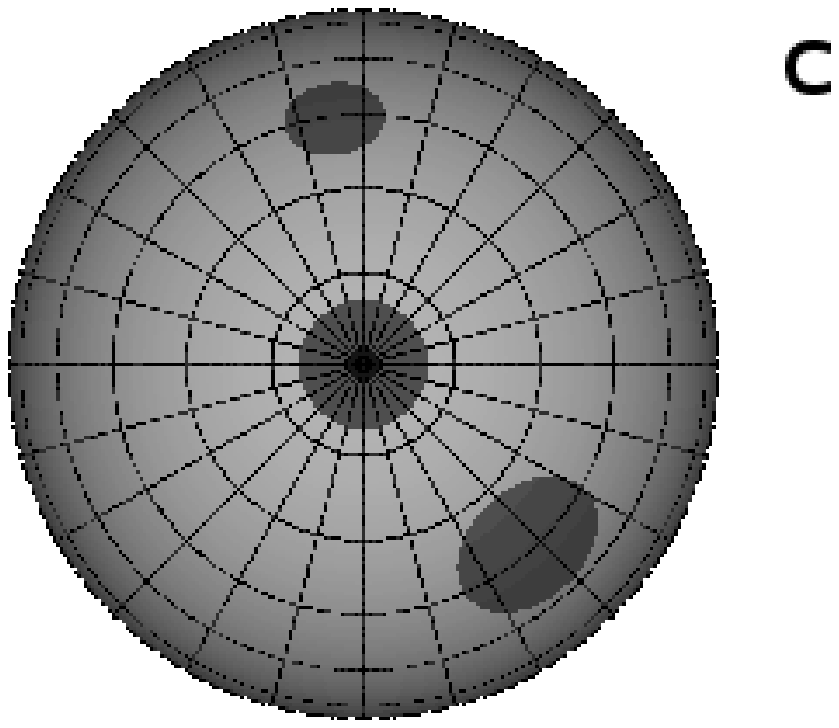}
	\includegraphics[width=41mm,bb=14 14 335 255]{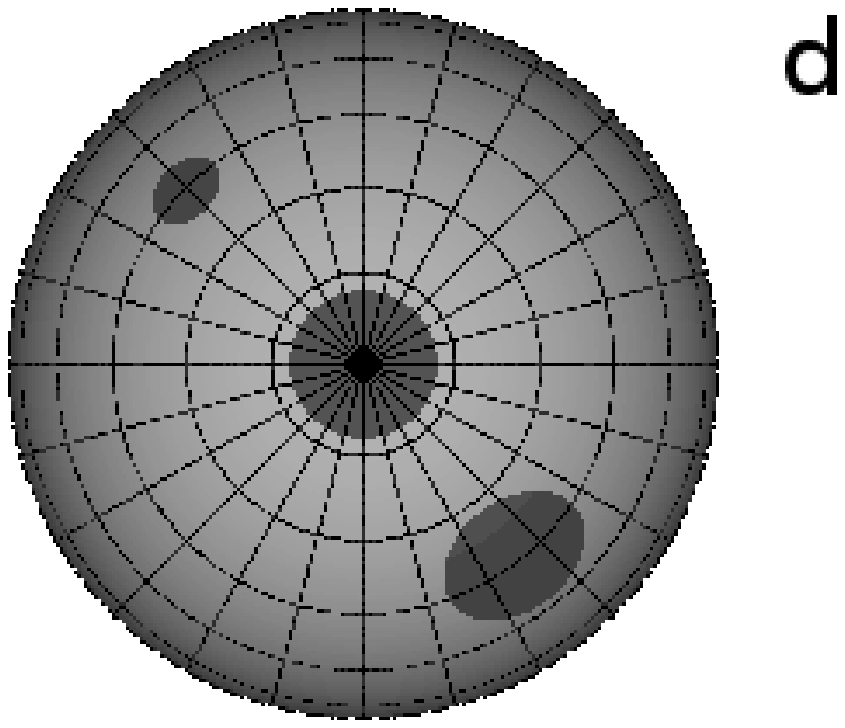}
	\caption{Results of the spot modelling for the labelled light curves in Figure \ref{fig:phases}. The models are displayed in the same phase in polar view.}
	\label{fig:models}
\end{figure}

The modelling was done for four parts of our observational data, between JDs
2453611--22,
2453846--67,
2454021--25 and
2454054--68. All these parts cover whole rotations. The phased light curves can be seen on Figure \ref{fig:phases}, whereas the results of the modelling are displayed on Figure \ref{fig:models}. One can easily see, that the shape of the light curve is continuously changing. During the time of the first season the light curve has an "S" shape. This indicates that only one large active area causes the modulation of the light curve, which gets in our line of sight at the phase of 0.4. When looking at the corresponding model: \textbf{(a)} on Figure \ref{tab:sml}, one can see that the modelling resulted in two relatively close stellar spots with an angular separation of about $100^\circ$. Because the fitting is done by assuming circular stellar spots, this could also indicate one large, longitudinally elongated active region present on the stellar surface. 

There is a large gap in our observations, and during that time considerable changes occurred on the stellar surface: on the next light curve \textbf{(b)} the shape of the light curve changes from an "S" shape to "W", which means that another active region is formed well separated from the previous one. The angular separation of the two non-polar spotted regions became larger by about $30^\circ$. The spot configuration remains the same on the next light curve \textbf{(c)}, but since the total brightness is increased (Figure \ref{fig:all_v}), the modelling results in smaller polar spots, which might reflect the longer activity cycle of EY Draconis. The separation between the active regions remains practically the same.
On the last light curve \textbf{(d)}, the "W"-shape of the light curve gets even more impressive. The minima are at $\sim$0.4 and $\sim$0.9 phase, so the spots are separated by $\sim$180 degrees. The spot modelling results are given in Table \ref{tab:sml}.

\begin{table}
\centering
 \begin{tabular}{r|c|c|c}
 \multicolumn{4}{c}{2453611--22 (a)}\\
\hline \hline
$\beta$ & 45 & 45 & 90 \\
$\lambda$ & $90\pm 2$ & $203\pm 3$ & 360 \\
$\gamma$ & $11.8\pm0.3$ & $10.2\pm0.3$ & $23.5\pm0.3$ \\
\hline
\end{tabular}
\begin{tabular}{r|c|c|c}
  \multicolumn{4}{c}{2453846--67 (b)}\\
\hline \hline
$\beta$ & 45 & 45 & 90 \\
$\lambda$ & $126\pm 3$ & $268\pm 4$ & 360 \\
$\gamma$ & $12.6\pm0.5$ & $10.3\pm0.5$ & $24.5\pm0.6$ \\
\hline
\end{tabular}

\begin{tabular}{r|c|c|c}
  \multicolumn{4}{c}{2454021--25 (c)}\\
\hline \hline
$\beta$ & 45 & 45 & 90 \\
$\lambda$ & $138\pm 3$ & $280\pm 9$ & 360 \\
$\gamma$ & $13.3\pm0.6$ & $8.6\pm1.0$ & $18.6\pm1.2$ \\
\hline
\end{tabular}

\begin{tabular}{r|c|c|c}
  \multicolumn{4}{c}{2454054--68 (d)}\\
\hline \hline
$\beta$ & 45 & 45 & 90 \\
$\lambda$ & $128\pm 3$ & $316\pm 10$ & 360 \\
$\gamma$ & $12.6\pm0.5$ & $6.1\pm1.0$ & $20.3\pm0.9$ \\
\hline
\end{tabular}

\caption{Results of spot modelling: the spot parameters in degrees. Those parameters, where no errors are given, are kept fixed during the modelling. $\beta$, $\lambda$ and $\gamma$ stand for the latitude, longitude and radius of the spot, respectively.}
\label{tab:sml}
\end{table}

\section{Stellar cycles}
\label{sect:analysis}
Fourier-analysis was done using MUFRAN (MUlti FRe\-quency ANalysis) tool made by \cite{mufran}. For the Budapest data we get a rotation period of $P_\mathrm{rot}=0^d.4587$, which is very close to those of \cite{Robb:IBVS95}. The rotational period seems to be stable during all the observations. 

\begin{figure}
	\centering
	\includegraphics[width=83mm]{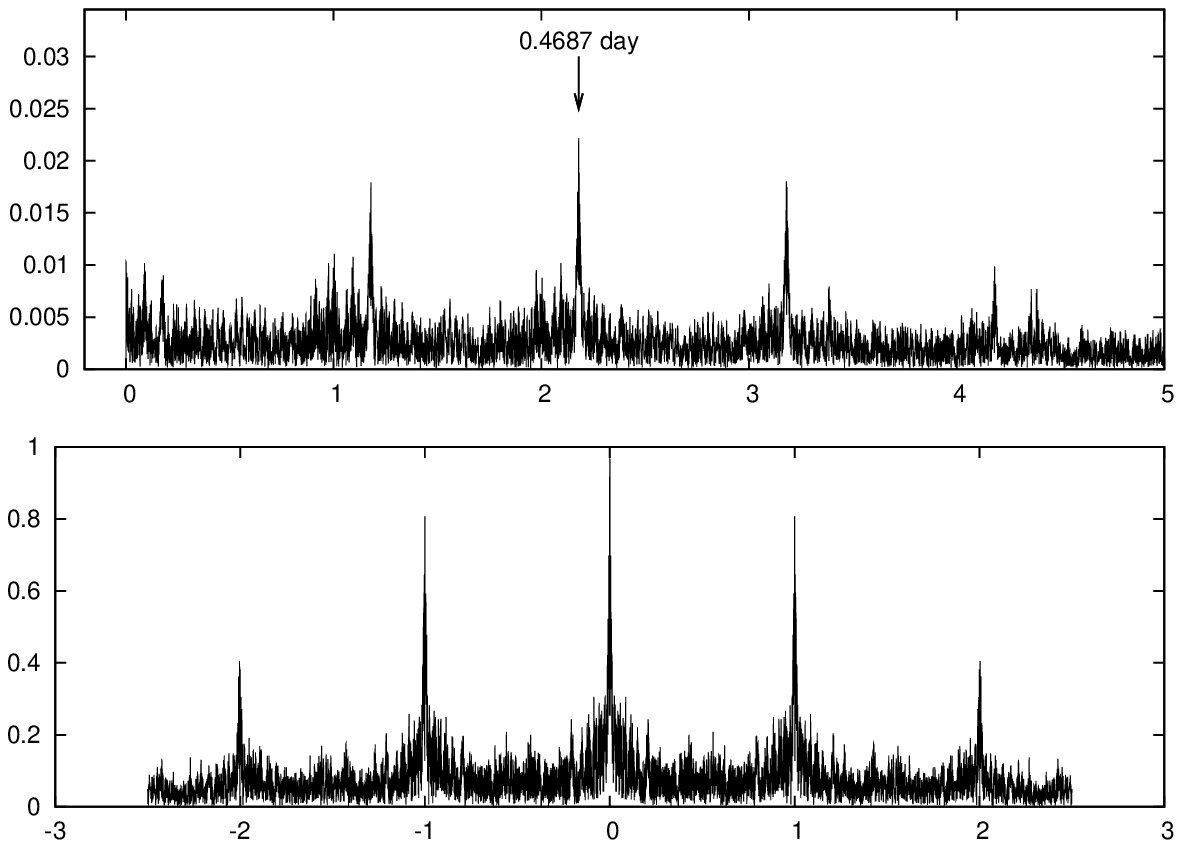}
	\includegraphics[width=83mm]{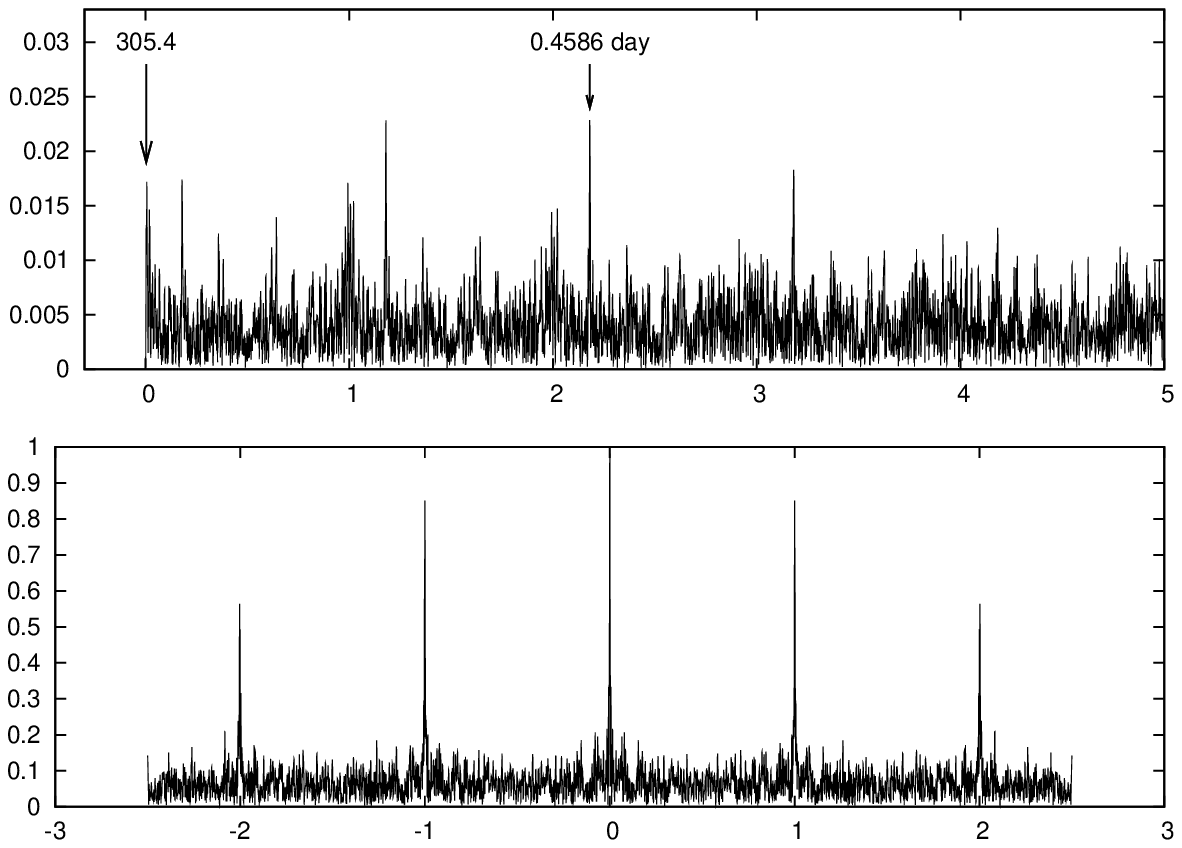}
	\caption{Fourier-spectrum and spectral window for the Budapest (top two panels) and ROTSE (lower two panels) light curve}
	\label{fig:sp_bp}
\end{figure}

Analysing the ROTSE data shows two frequencies: one corresponding to the rotational period (this one is the same as in Budapest data), and the other one suggests a longer period of 300 days.  This longer cycle cannot be detected in the Budapest observations, since there is a 200 days-long gap in the observations. The Fourier-spectrum and the corresponding spectral window of the Budapest observations can be seen on Figure \ref{fig:sp_bp}. The brightness increase between JDs 2453939 and 2453953 however seems to confirm this longer cycle, but this suggestion should be handled with care: the first reason is, that only one season of photometry suggests the presence of this cycle. The second is, that the total length of our observation is less than 500 days, which is not much longer that the cycle we would like to prove.

If this activity cycle would be confirmed, EY Draconis would be placed on the rotational period--activity cycle diagram of \cite{cycles} as the first ultrafast rotator with known very short activity cycle.

\begin{figure}
	\centering
	\includegraphics[width=83mm]{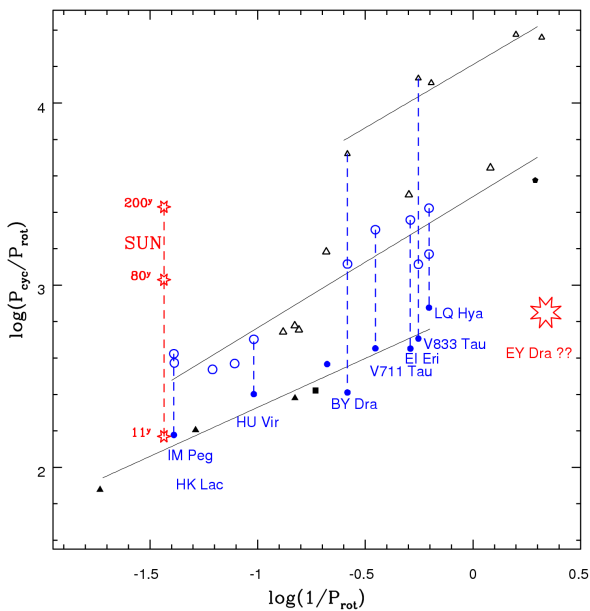}
	\caption{Relation between the rotational period and stellar cycles after \cite{cycles}.}
	\label{fig:cycles}
\end{figure}
\section{Conclusions}
\begin{itemize}
\item On short timescale the light curve of EY Draconis is found to be stable. This stability could be caused by large scale stable magnetic fields, or by an unresolvable binary component.
 \item The changes of the light curve shape indicates the spot evolution on the stellar surface. The initial spot configuration of one elongated active region (or two close stellar spots) changes to two detached spots, which is resolved by spot modelling. The evolution of the stellar surface can be continuously followed both in the changing shape of the light curve and by the spot models.
\item The size of the polar spot changes parallel with the maximum brightness of EY Draconis, which could reflect the activity cycle: as the polar spot gets larger, the maximum brightness of the lightcurve decreases. 
\item When analysing the light curve obtained from the ROTSE database, beside the rotational modulation another period can be found in the Fourier-spectrum, which could indicate the presence of a longer, 300-day stellar cycle. The large, 200-days long gap in our observations prevents us to detect the cycle.
\item If the presence of the stellar cycle is confirmed, this star would be the first ultra-fast rotator with short spot cycle on the plot of \cite{cycles} which would strengthen the relation found in the paper of \cite{cycles}.
\end{itemize}

\acknowledgement{
We would like to thank K. Ol\'ah, J. Jurcsik, B. Szeidl, I. D\'ek\'any, Zs. S. Hurta, K. Posztob\'anyi, \'A. S\'odor, A. \hbox{Szing}, M. V\'aradi for their help in various parts of this work, and for making a lot of the observations. K. V. appreciates the hospitality of Konkoly Observatory. 
The financial support of the Hungarian government through OTKA T043504 and T048961 is acknowledged.
}

\end{document}